\documentclass{vgtc}                          

\ifpdf
  \pdfoutput=1\relax                   
  \usepackage{graphicx}                
  \DeclareGraphicsExtensions{.pdf,.png,.jpg,.jpeg} 
\else
  \usepackage{graphicx}                
  \DeclareGraphicsExtensions{.eps}     
\fi%

\graphicspath{{figures/}{pictures/}{images/}{./}} 

\usepackage{microtype}                 
\PassOptionsToPackage{warn}{textcomp}  
\usepackage{textcomp}                  
\usepackage{mathptmx}                  
\usepackage{times}                     
\usepackage{cite}                      
\usepackage{tabu}                      
\usepackage{booktabs}                  

\usepackage{caption}
\usepackage{subcaption}
\usepackage{enumerate}

\usepackage{hyperref}

\usepackage{footnote}

\onlineid{0}

\vgtccategory{Research}

\title{VAINE: Visualization and AI for Natural Experiments}

\author{Grace Guo\thanks{e-mail: gguo31@gatech.edu}\\ %
    \scriptsize Georgia Institute of Technology %
\and Maria Glenski\thanks{e-mail: maria.glenski@pnnl.gov}\\ %
     \scriptsize Pacific Northwest National Laboratories %
\and ZhuanYi Shaw\thanks{e-mail: yi.shaw@pnnl.gov}\\ %
     \scriptsize Pacific Northwest National Laboratories %
\and Emily Saldanha\thanks{e-mail: emily.saldanha@pnnl.gov}\\ %
     \scriptsize Pacific Northwest National Laboratories
\and Alex Endert\thanks{e-mail: endert@gatech.edu}\\ %
     \scriptsize Georgia Institute of Technology
\and Svitlana Volkova\thanks{e-mail: svitlana.volkova@pnnl.gov}\\ %
     \scriptsize Pacific Northwest National Laboratories
\and Dustin Arendt\thanks{e-mail: dustin.arendt@pnnl.gov}\\ %
     \scriptsize Pacific Northwest National Laboratories
     }

\teaser{
  \centering
  \includegraphics[width=\linewidth]{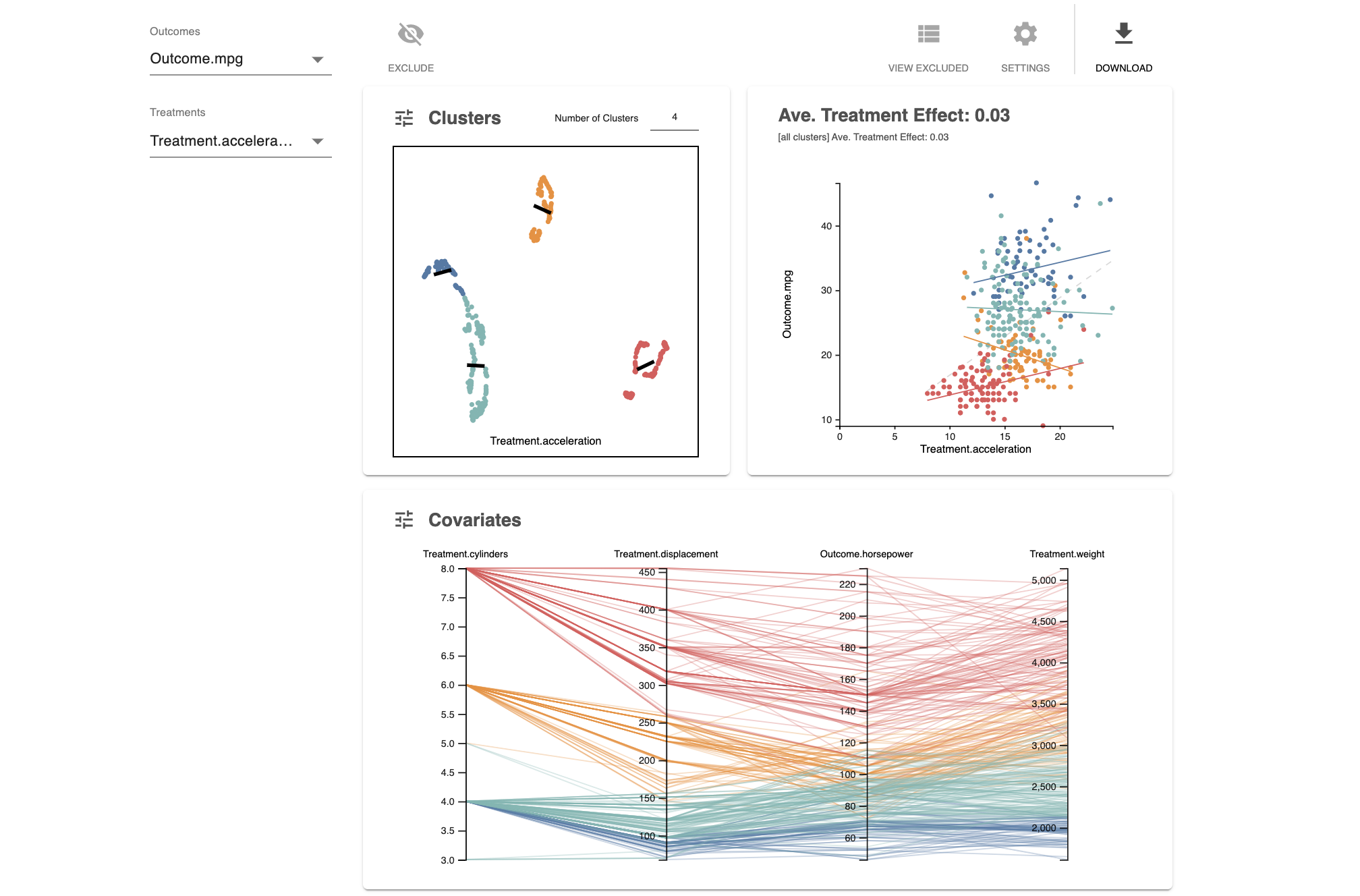}
  \caption{The VAINE (Visualization and AI for Natural Experiments) system interface. The three main views --- the \textit{Clusters} view, the \textit{Average Treatment Effect} view, and the \textit{Covariates} view --- enable analysts to estimate the effect size of the treatment on the outcome variable of interest. Interactive menu options can be used to adjust clusters, exclude outliers, and inspect covariates.
}
  \label{fig:teaser}
}

\abstract{Natural experiments are observational studies where the assignment of treatment conditions to different populations occurs by chance ``in the wild''. Researchers from fields such as economics, healthcare, and the social sciences leverage natural experiments to conduct hypothesis testing and causal effect estimation for treatment and outcome variables that would otherwise be costly, infeasible, or unethical.
In this paper, we introduce VAINE (Visualization and AI for Natural Experiments), a visual analytics tool for identifying and understanding natural experiments from observational data.
We then demonstrate how VAINE can be used to validate causal relationships, estimate average treatment effects, and identify statistical phenomena such as Simpson’s paradox through two usage scenarios.} 

\CCScatlist{
  \CCScatTwelve{Human-centered computing}{Visu\-al\-iza\-tion}{Visualization systems and tools }{};
  \CCScatTwelve{Human-centered computing}{Visu\-al\-iza\-tion}{Visualization application domains}{Visual analytics}
}

\begin{document}


\firstsection{Introduction}

\maketitle

Identifying and quantifying causal relationships are essential tasks in many domains to develop theories, build models, and guide policy.
Accurate causal inference depends on running controlled experiments to investigate the variables in question. 
This requires the random allocation of treatment conditions via randomized controlled trials (RCTs).
RCTs allow scientists to account for confounding factors, and are considered the ``gold standard'' for supporting causal claims.
%
However, RCTs may not be possible, especially in situations where manipulating treatment variables is impractical or unethical. 

For example, during the COVID-19 pandemic, understanding causal relationships between non-pharmaceutical interventions and infection transmission would have helped policy makers decide which interventions to implement and when. However, it is both unethical and impractical to conduct RCTs by dictating different health guidelines to arbitrary subsets of populations.
Researchers must thus rely on observational data available post-hoc to look for indirect clues about causality.
%
Beyond global pandemics, similar approaches are needed to understand the causal relationships that explain human behavior, information operations (what \textit{causes} successful disinformation campaigns to go viral online?), and model evaluations (what \textit{causes} a model to fail?).
%


Causal claims can be tested using observational data if a \textit{natural experiment} has occurred.
This is an uncontrolled process where the assignment of treatment conditions to different groups has occurred ``as if at random''. 
Such natural experiments have long been used by researchers in fields such as economics, healthcare, and the social sciences.
By analyzing observational data, researchers are able to conduct hypothesis testing and causal effect estimation that would otherwise be costly, infeasible, or unethical \cite{dinardo2010natural}.



There are many challenges and caveats to working with observational data.
In particular, the well known adage that \textit{correlation doesn't imply causation} duly applies.
Correlations between treatment and outcome variables may be driven by underlying confounding variables that aren't observed.
Making a claim of causation thus requires showing that correlation still exists when controlling for all other variables, i.e., \textit{covariates}.
As such, while algorithmically generated claims of causation can be used to explore potential causalities in observational data, they shouldn't be taken at face value and need to be interpreted and contextualized by a domain expert.
When working with natural experiment data, experts would want to know whether all possible covariates have been accounted for during data collection,
how much outliers influence the result,
and what characterizes subgroups experiencing stronger effects.
The need to contextualize, explore, and interpret the results of natural experiments thus motivates the development of a visual analytics tool to better support human-in-the-loop causal analysis.



In this paper, we introduce VAINE (Visualization and AI for Natural Experiments) (Fig. \ref{fig:teaser}), a visual analytics system that enables users to identify natural experiments in observational data and estimate the impact of treatment variables on outcomes. 
We also demonstrate how VAINE can be used to validate causal relationships, estimate average treatment effects, and identify statistical phenomena such as Simpson’s paradox through two usage scenarios.

\section{Related Work}

Causal inference from observational data has long been a challenge in statistics and data analysis.
Much work has been done in the field of mathematical causal discovery \cite{pearl1995causal, mani2004causal} and causal structure analysis \cite{nadkarni2001bayesian}.
More recently, a number of visualization techniques have been developed to support causality representation \cite{elmqvist2003causality, bae2017understanding} and inference \cite{wang2015visual, xie2020visual}.
However, causality inference from observational data remains a challenging task, particularly because \textit{correlation does not imply causation}. For instance, a study by Shalizi et al. on social network data has found that the use of regression coefficients, even taking into consideration temporal data, cannot distinguish causal factors from confounding variables \cite{shalizi2011homophily}.


Despite the difficulties of casual inference, once a causal relationship has been identified, it becomes possible to conduct hypothesis testing by identifying natural experiments when they occur in observational data.
Natural experiments have long been used by researchers in fields such as economics \cite{angrist1990lifetime, jensen2007digital}, healthcare \cite{finkelstein2012oregon} and the social sciences \cite{angrist1999using}.
Observational data can thus be successfully used to identify natural experiments when they occur, allowing analysts to estimate the effect size of the treatment variable on the outcome variable of interest.
However, natural experiments remain an underexplored area for visual analytics systems. In this paper, we aim to address this gap through the VAINE system.
%

\section{VAINE}


During the design and implementation of VAINE, the first author was embedded within a research team of domain experts applying natural experiments to observational data.
In addition to causal effect estimation, the results had to be interpreted and contextualized.
Based on these requirements and prior related work, the following design goals (DG) were derived collaboratively:

\begin{enumerate}
  \item[\textbf{DG1:}] \textbf{Control for confounding variables.} VAINE should control for confounding variables by automatically grouping data items such that items in the same group have similar values for all covariates other than the treatment and outcome variables.
  \item[\textbf{DG2:}] \textbf{Estimate effect sizes.} The primary outcome of VAINE should be an estimate of the effect of the treatment on the outcome variables. Users should be able to obtain this estimate, and understand how it is derived from the data clusters.
  \item[\textbf{DG3:}] \textbf{Incorporate user feedback.} Users should be able to incorporate domain expertise or contextual information to guide the algorithm. For example, by adjusting clustering parameters, removing outliers, and inspecting cluster details.
  \item[\textbf{DG4:}] \textbf{Support responsive UI.} 
Interactions should automatically trigger rapid recomputation of analytics and update visualizations.
\end{enumerate}

VAINE \footnote{\url{https://github.com/pnnl/vaine-widget}}
is a domain-agnostic tool implemented as a python package and widget for the popular Jupyter notebook \footnote{\url{https://jupyter.org}} computational environment. Our target users are research analysts who are familiar with using such computational environments for causal analysis.
The front-end is built in JavaScript, using the React \footnote{\url{https://reactjs.org}} framework.




\subsection{Interface} \label{interface}
The VAINE system displays the current treatment and outcome variables in the far left panel. The interface has three main views: \textit{Clusters}, \textit{Average Treatment Effect}, and \textit{Covariates} views (Fig. \ref{fig:teaser}). 
The same cluster color coding is used to coordinate across views. Clusters with an insignificant ($p > 0.05$) regression coefficient are deselected by default (grayed out) across all views.

The \textit{Clusters} view displays all data instances clustered based on their covariates, with a default of 10 clusters.
Data instances in the same cluster have similar covariate values, and only vary by treatment and outcome values.
This controls for confounding variables, and isolates the effect of the treatment on the outcome variable \textbf{(DG1)}.
Each cluster also has an overlaid line indicating the linear regression of the outcome variable on the treatment variable for all data instances in that cluster \textbf{(DG2)}. Note that this regression line is based on the data values of the treatment and outcome variables, not the dimensionality reduced coordinates of data points.

The \textit{Average Treatment Effect} view displays data points in a linear model plot based on their treatment and outcome values. Each cluster is overlaid with a line of the linear regression of the outcome variable on the treatment variable. 
This view includes the weighted average treatment effect for selected clusters, an estimate of the impact of the treatment on the outcome variable \textbf{(DG2)}.
The overall regression line for the entire dataset (without clustering) is depicted with a faint dashed line, providing context to interpret subgroup regressions relative to that of the entire dataset. 


The \textit{Covariates} view summarizes covariate values of all data instances in a parallel coordinates plot, with only the first five covariates shown by default. Visual clustering is used to enhance patterns in the data and reduce visual clutter \cite{zhou2008visual}. In this view, analysts can explore how the clustering controls for covariate values \textbf{(DG1)}.


\subsection{Interactions} \label{interactions}
Treatment and outcome variables of interest can be selected using the drop-down menus on the left.
VAINE then automatically treats all other variables in the dataset as covariates, and re-clusters data instances based on the new covariates \textbf{(DG4)}.

In the \textit{Clusters} view,
analysts can manually adjust selected clusters using a drop-down menu, or by clicking on their regression line overlay \textbf{(DG3)}.
In order to coordinate across views, 
cluster selections propagate to all three views.
Additionally, the \textit{Clusters} view also allows analysts to adjust the number of clusters created \textbf{(DG3)}. When this value is changed, VAINE automatically re-clusters all data instances and updates all three views.
Clicking on a cluster name in the drop-down menu opens a dialogue box where users can change cluster name and color, and inspect the distribution of its covariates \textbf{(DG1)}.
In the \textit{Covariates} view, analysts can manually select which covariates to inspect using the drop-down menu \textbf{(DG1)}. This allows users to prioritize covariates based on their relative importance.

Brushing and linking is used to coordinate analysis across all views.
Users can brush over any of the views 
to select a subset of data points.
Selected points, such as outliers, can be excluded from analysis using the \textit{Exclude} button.



\subsection{Algorithm}
During preprocessing, the system
computes a 2-D embedding using UMAP\cite{mcinnes2018umap} for each treatment with the covariates and remaining treatments as input features \textbf{(DG1)}.
The tool then performs hierarchical clustering of each 2-D embedding and outputs a dendrogram that is passed to the system front-end using the approach described by Arendt et al.\cite{arendt2020parallel}.
On load, the number of clusters defaults to 10.
When the number of clusters is adjusted, the dendrogram from the preprocessing phase is used to re-cluster the $n$ data points in $O(n)$ time, allowing VAINE to quickly respond to interactions \textbf{(DG4)}.

The average treatment effect ($ATE$) value estimates the size of the impact of the treatment on the outcome variable, and allows analysts to perform causal effect estimation.
The $ATE$ is calculated using the weighted average of the regression coefficients of all selected clusters \textbf{(DG2)}.
In order to do so, linear regression of the outcome variable on the treatment variable is first performed for all clusters.
Linear regression was chosen as the default algorithm because it is simple and can be quickly calculated on the front end, allowing for rapid user adjustments \textbf{(DG4)}.
For a dataset with $M$ selected clusters and a sum of $N$ data instances across these clusters, each cluster has a size of $n_{i}$ (such that $n_1 + n_2 + ... + n_{M} = N$) and a regression coefficient of $b_{i}$. The calculated $ATE$ of this dataset is:

\begin{equation}
    ATE = \frac{1}{N} \sum_{i=1}^{M}{n_i \cdot b_i}
\end{equation}

When analysts exclude data points from the dataset (described in \ref{interactions}), the clusters will not change. However, regression equations will be recalculated for affected clusters, and the $ATE$ will be updated using the new regression coefficients and cluster sizes \textbf{(DG4)}.

\section{Usage Scenarios}
We demonstrate VAINE with two usage scenarios \cite{isenberg2013systematic}. The first uses the Auto MPG dataset \cite{Dua:2019}, often used to predict the miles per gallon of cars from variables such as weight. The second scenario looks at the Ames Housing dataset \cite{de2011ames}, which has been used to predict house sale prices from other variables.
These datasets were selected because they are intuitive and contain well-established causal relationships, which allow us to validate VAINE.

\subsection{Usage Scenario 1: Auto MPG Dataset}

The Auto MPG dataset \cite{Dua:2019} contains 9 attributes. 
We exclude temporal and ordinal variables, and consider \textit{mpg} and \textit{horsepower} to be the outcome variables.
All other variables are treatment variables.

\begin{figure}[tb]
     \centering
     \begin{subfigure}[tb]{\columnwidth}
         \centering
         \includegraphics[width=\columnwidth]{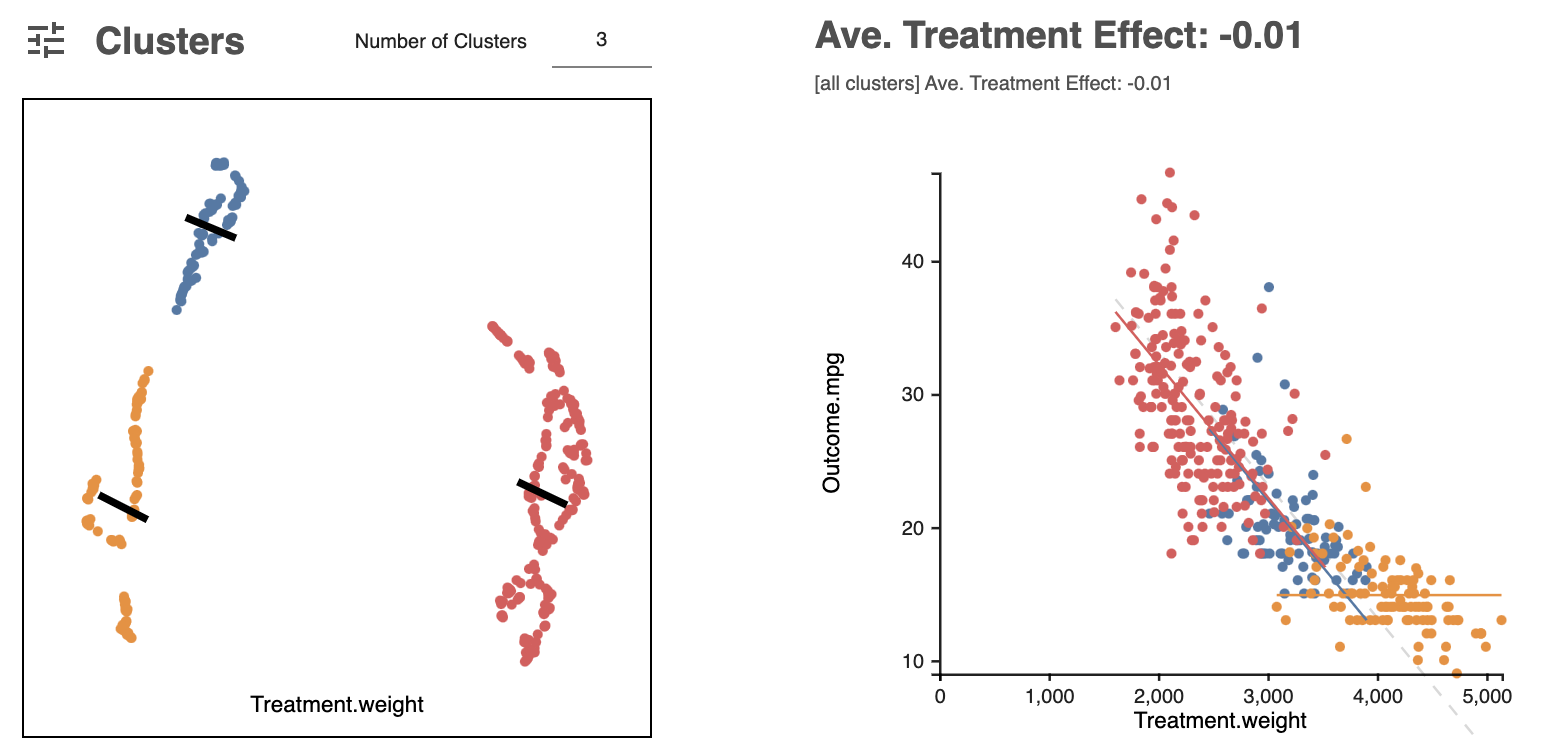}
         \caption{From the \textit{Clusters} view, we identify 3 distinct clumps in the dataset. After setting the number of clusters to 3, the \textit{ATE} of \textit{weight} on \textit{mpg} is estimated to be -0.01.}
         \label{fig:demo_cars_1}
     \end{subfigure}
     \hfill
     \begin{subfigure}[tb]{\columnwidth}
         \centering
         \includegraphics[width=\columnwidth]{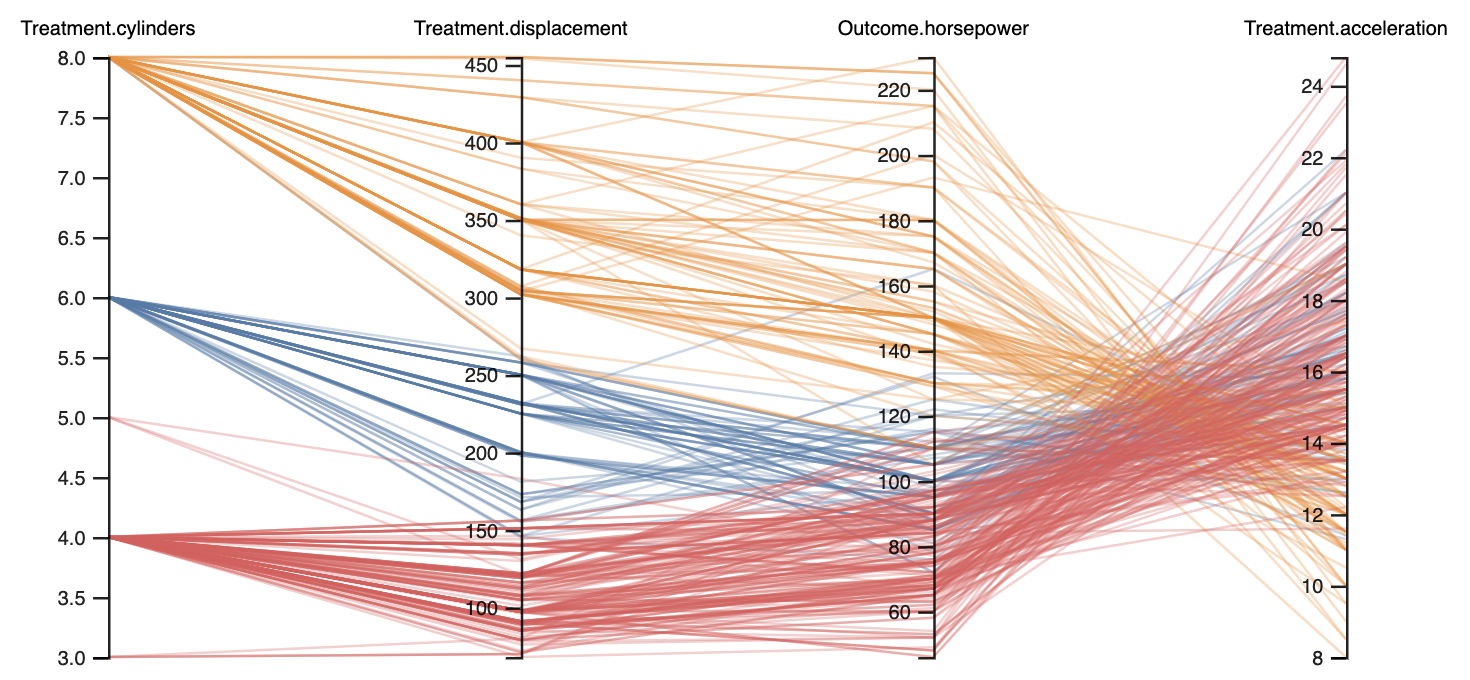}
         \caption{From inspecting the \textit{Covariates} view, we can see that the three clusters are clearly distinguished by their covariates. In particular, instances within the same cluster share similar values for the number of \textit{cylinders}, \textit{displacement} and \textit{horsepower}.}
         \label{fig:demo_cars_2}
     \end{subfigure}
        \caption{
        After setting the number of clusters for a treatment and outcome pair, the clustering can be verified in the \textit{Covariates} view.}
        \label{fig:demo_cars}
\end{figure}


It has often been claimed that the weight of a vehicle affects its fuel economy \cite{epa_2021} (i.e., its miles per gallon (\textit{mpg}) value). We can look at this proposed causal relationship in VAINE by selecting \textit{weight} as the treatment variable of interest.
In the \textit{Clusters} view, we see three distinct clumps in the dataset. Setting the number of clusters to 3 identifies these (Fig. \ref{fig:demo_cars_1}).
The \textit{Covariates} view reveals that the clusters are clearly distinguished by their covariates. In particular, instances within the same cluster share similar values for the number of \textit{cylinders}, \textit{displacement} and \textit{horsepower} (Fig. \ref{fig:demo_cars_2}).
From the \textit{Average Treatment Effect} view, we can see a small but significant estimated effect of \textit{weight} on \textit{mpg}.
VAINE thus allows us to confirm from the data that when other variables are held nearly constant, increasing a car's weight decreases its \textit{mpg},
a finding that agrees with both EPA guidelines and prior studies such as \cite{wang2015visual}. 




For comparison, we can also look at variables not known to affect \textit{mpg}, such as \textit{acceleration}.
A simple linear regression of \textit{acceleration} on \textit{mpg} for the entire dataset yields a regression coefficient of 1.20.
We can verify this relationship in VAINE by selecting \textit{acceleration} as the treatment variable of interest.
The \textit{Clusters} view reveals multiple distinct clusters in the dataset; we thus adjust the number of clusters to 4 (Fig. \ref{fig:teaser}).
With this new clustering, the $ATE$ is now 0.03.
While still significant, this is a large decrease from the  coefficient of 1.20 when only a simple linear regression was used without clustering. 
Additionally, the direction of the effect of \textit{acceleration} on \textit{mpg} is not constant across clusters.
The overall regression line for the entire dataset suggests a strongly positive relationship of \textit{acceleration} on \textit{mpg} (Fig. \ref{fig:teaser}).
However, from examining the subgroups, we see that the blue and orange clusters show a relationship opposite to the overall trend.
This is a case of Simpson's paradox, where trends in subgroups disappear or are reversed when groups are combined.
By visualizing clusters in the dataset, VAINE thus supports the identification of inconsistent treatment effects in subgroups, allowing instances of Simpson's paradox to be mitigated by expert judgement.



\subsection{Usage Scenario 2: Ames Housing Dataset}
The Ames Housing dataset \cite{de2011ames} contains 81 attributes. We exclude all temporal and ordinal attributes,
and consider \textit{Sale Price} to be the outcome attribute of interest. All other attributes are treatments.

\begin{figure}[tb]
 \centering
 \includegraphics[width=\columnwidth]{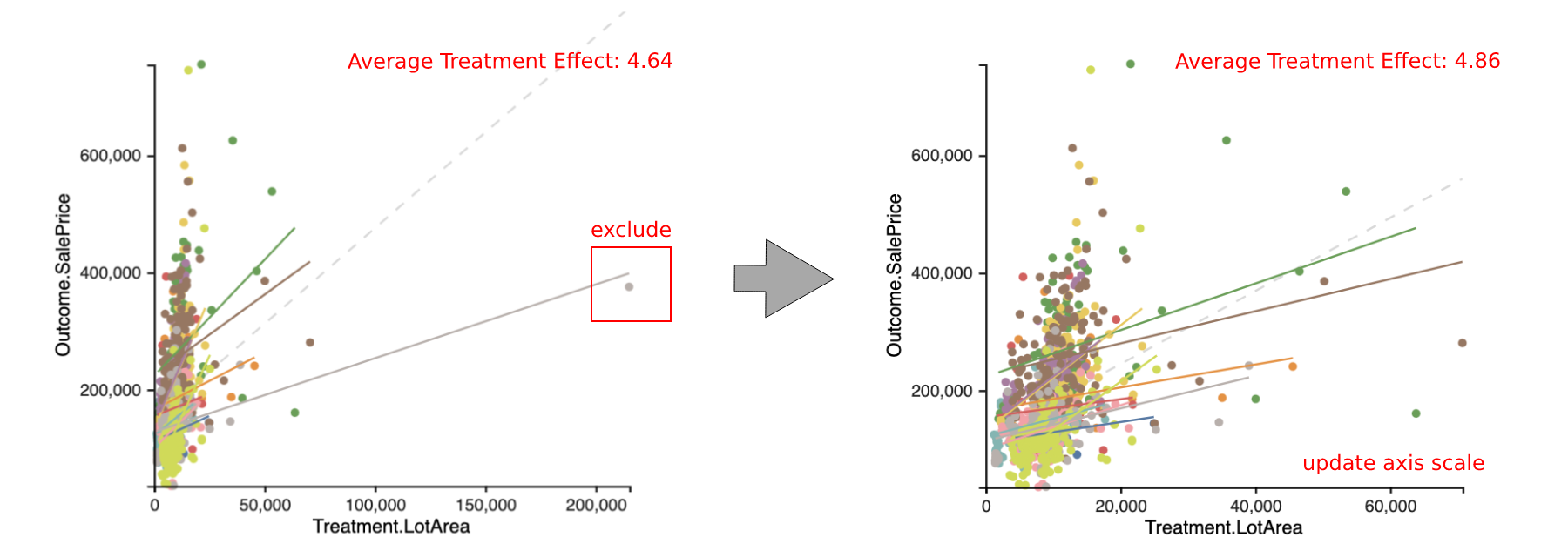}
 \caption{Outliers identified in the \textit{Average Treatment Effect} view can be excluded from analysis, updating the axes and $ATE$ value.
}
 \label{fig:demo_houses_1}
\end{figure}

Of the available treatment variables, we would expect property area to affect property sale price.
In order to estimate the change in sale price per unit change in property area, we set \textit{Lot Area} as the treatment variable. From the \textit{Average Treatment Effect} view, a clear outlier is identified.
Excluding this outlier increases the estimated $ATE$ of \textit{Lot Area} on \textit{Sale Price} to 4.86 (Fig. \ref{fig:demo_houses_1}).

\begin{figure}[tb]
     \centering
     \begin{subfigure}[tb]{\columnwidth}
         \centering
         \includegraphics[width=\columnwidth]{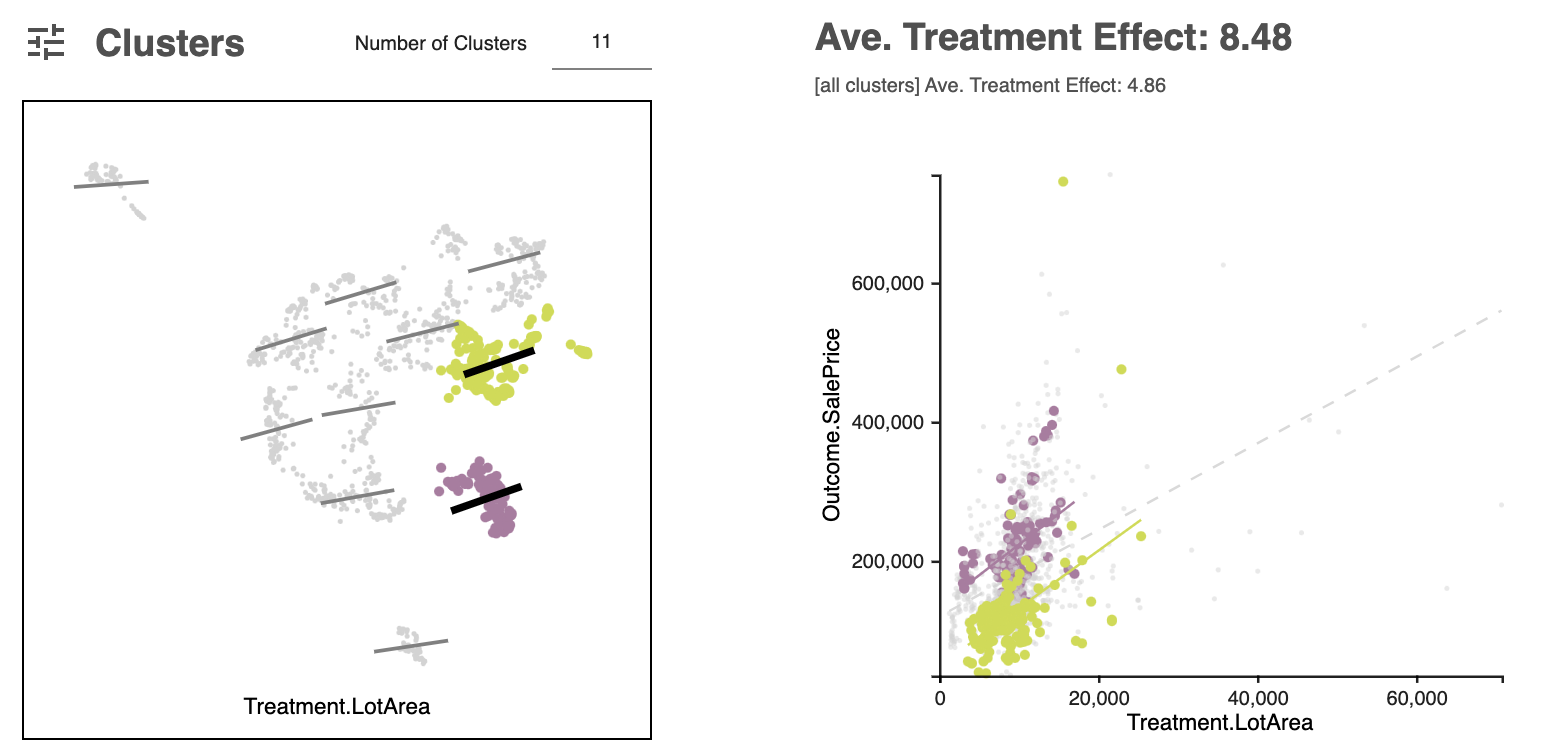}
         \caption{The purple and lime clusters have the steepest regression lines. Selecting these two clusters increases the $ATE$ to 8.48 (from an $ATE$ of 4.86 when all clusters are selected).}
         \label{fig:demo_houses_3}
     \end{subfigure}
     \hfill
     \begin{subfigure}[tb]{\columnwidth}
         \centering
         \includegraphics[width=\columnwidth]{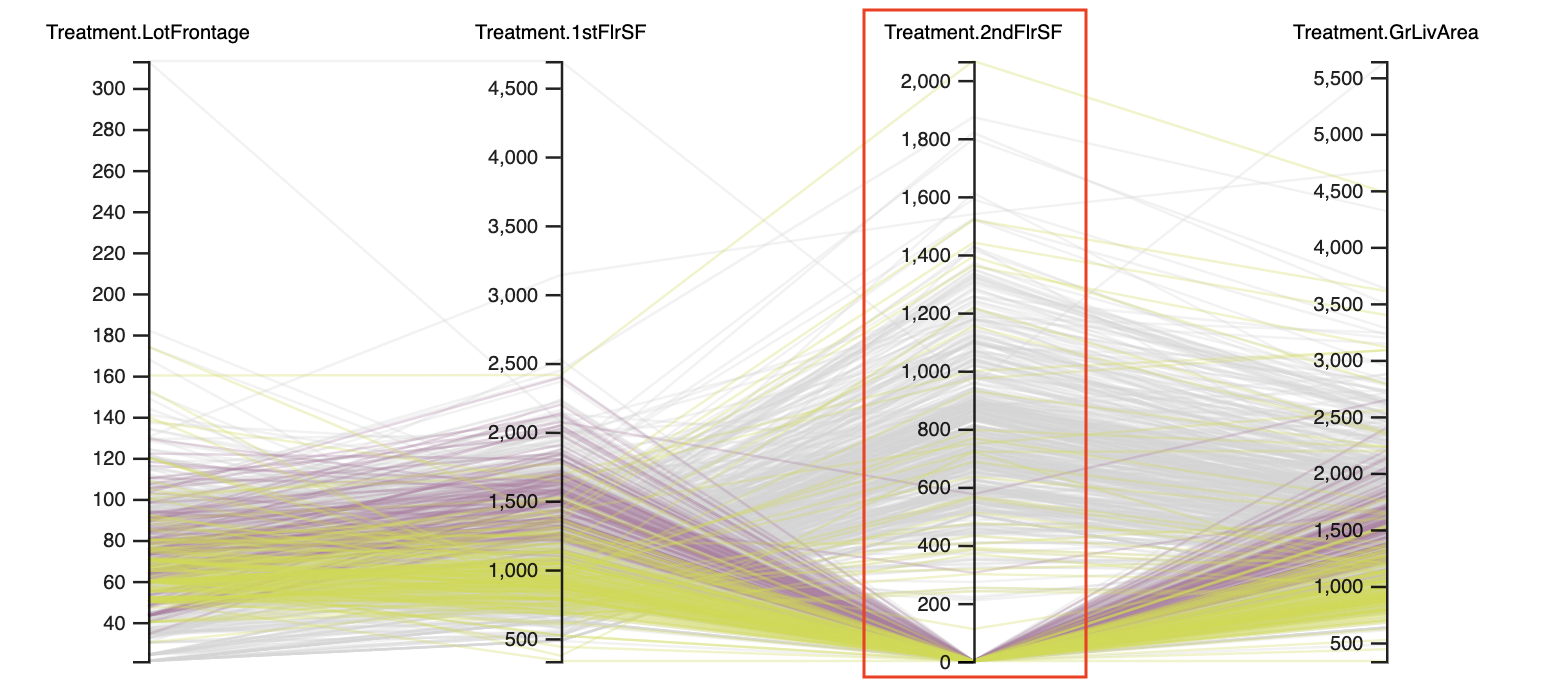}
         \caption{Unlike the rest of the dataset, data instances in the selected clusters (purple and lime) tend to have a \textit{2ndFlrSF} of 0. For clarity, only a subset of covariates are shown.}
         \label{fig:demo_houses_2}
     \end{subfigure}
        \caption{By inspecting the purple and lime clusters in greater detail, we can conclude that \textit{Lot Area} has a stronger effect on \textit{Sale Price} for smaller properties without a second floor.}
        \label{fig:demo_houses}
\end{figure}

In the \textit{Clusters} view, some clusters have a steeper regression line than others, suggesting that \textit{Lot Area} has a stronger effect on \textit{Sale Price} for those particular properties. We can investigate this further by deselecting all other clusters, highlighting only the purple and lime clusters with the steepest regression lines. From this selection, we see that the $ATE$ for these two clusters is 8.48, indicating that the effect of \textit{Lot Area} on \textit{Sale Price} for properties in these two clusters is almost double that of the entire dataset (Fig. \ref{fig:demo_houses_3}). From the \textit{Average Treatment Effect} view, we see that the two clusters tend towards smaller \textit{Lot Area} values. The steeper effect of \textit{Lot Area} on \textit{Sale Price} for smaller properties corresponds with findings from prior studies using the Ames housing dataset \cite{hohman2019gamut}.

Using the \textit{Covariates} view, we can further inspect the 
data instances in the two clusters.
From Fig. \ref{fig:demo_houses_2}, it can be seen that the most distinct covariate differentiating these clusters from the rest of the data is the Second floor square feet (\textit{2ndFlrSF}) variable. Instances in the two clusters tend to have a \textit{2ndFlrSF} of 0. We thus conclude that per unit increases in \textit{Lot Area} cause a greater increase in \textit{Sale Price} for smaller properties without a second floor.

\section{Limitations and Future Work} \label{limitations}
VAINE was originally designed to operate on continuous treatments, outcomes, and covariates. Therefore, some decisions are less appropriate for ordinal or categorical data, such as using dimension reduction and linear regression. However, workarounds such as one-hot encodings are possible.
We also note that VAINE assumes linearity when estimating the $ATE$. While the \textit{Average Treatment Effect} view can be used to identify instances of non-linear relationships, VAINE does not currently take such factors into consideration.
Moving forward, we would like to extend VAINE to account for non-linearity. Given the utility of time ordering to establish causal relationships, future work can also extend VAINE to account for a range of data types and experimental setups.



Finally, the design goals outlined in this paper were identified by the research team of domain experts conducting causal analysis. There may be further design goals not considered here.
Preliminary applications of VAINE by the research team have also found limitations in terms of the scalability of the system to datasets with tens of thousands of instances. While computation time remains feasible, optimizing for scale represents a potential avenue of future work.

\section{Conclusion}
In this paper we presented VAINE, a visual analytics tool designed for research analysts familiar with using computational environments for causal analysis.
VAINE helps users identify natural experiments in observational data
by clustering data instances based on covariates. Analysts can thus control for confounding variables and isolate the effect of the treatment(s) on outcome variables.
The visual representations in VAINE allow domain experts to explore potential causalities and contextualize results by adjusting clusters, identifying outliers, and inspecting covariates.
We have also demonstrated how VAINE can validate causal relationships, estimate average treatment effects, and identify cases of Simpson's paradox.

Despite the limitations addressed above, VAINE is a practical and novel technique for leveraging natural experiments in observational data for human-in-the-loop causal effect estimation.
Having validated VAINE, the research team is currently applying it to understand various operational domains including information operations, human behavior, and causal explanations of model performance.



\acknowledgments{
The authors wish to thank the anonymous reviewers for their thoughtful and detailed feedback on this paper, and members of the Georgia Tech Visualization Lab for their helpful insights and comments. 

This research was developed with funding from the Defense Advanced Research Projects Agency (DARPA). The views, opinions and/or findings expressed are those of the author and should not be interpreted as representing the official views or policies of the Department of Defense or the U.S. Government. Further development and evaluation  of this work is supported by NSF IIS-1750474.
}

\bibliographystyle{abbrv-doi}

\bibliography{template}
\end{document}